\documentclass[preprint,
aps,amsmath,byrevtex,prd,titlepage,
preprintnumbers,
tightenlines,
nofootinbib]{revtex4-1}

\usepackage{amsmath}
\usepackage{amssymb}
\usepackage{bm}
\usepackage{hyperref}
\usepackage{graphicx}
\usepackage{slashed}
\usepackage{dcolumn}

\newcommand{\beq}{\begin{equation}}
\newcommand{\eeq}{\end{equation}}
\newcommand{\eq}[1]{Eq.~(\ref{#1})}

\begin{document}

\title {Hyperfine Splitting in Muonium: Accuracy of the Theoretical Prediction}
\author {Michael I. Eides}
\altaffiliation[Also at ]{the Petersburg Nuclear Physics Institute,
Gatchina, St.Petersburg 188300, Russia}
\email[Email address: ]{meides@g.uky.edu, eides@thd.pnpi.spb.ru}
\affiliation{Department of Physics and Astronomy,
University of Kentucky, Lexington, KY 40506, USA}

\begin{abstract}

In the last twenty years, the theory of hyperfine splitting in muonium developed without any experimental input. Finally, this situation is changing and a new experiment on measuring hyperfine splitting  in muonium is now in progress at J-PARC. The goal of the MuSEUM experiment is to improve by an order of magnitude experimental accuracy of the hyperfine splitting and muon-electron mass ratio. Uncertainty of the theoretical prediction for hyperfine splitting will be crucial for comparison between the forthcoming experimental data and the theory in search of a possible new physics. In the current literature estimates of the error bars of the theoretical prediction differ roughly by a factor of two.  We explain the origin of this discrepancy and obtain the theoretical prediction for the muonium hyperfine splitting $\Delta \nu^{th}_{\scriptscriptstyle HFS}(Mu)=4~463~302~872~(515)~\mbox{Hz},\; \delta=1.2\times 10^{-7}$.

\end{abstract}

\maketitle

\section{Introduction}

Calculations of hyperfine splitting (HFS) in one-electron atoms have a long and distinguished history starting with the classic works by Fermi \cite{f1930} and Breit \cite{br1931}. The modern state of the HFS theory in muonium was reviewed in every detail in \cite{egs2001,egs2007}. Small corrections to HFS calculated after publication of these reviews are collected in
\cite{codata2014}. High precision measurements of HFS in muonium  for a long time were considered as a test of the high precision QED and a source for precise values of the fine structure constant $\alpha$ and the muon-electron mass ratio $m_\mu/m_e$. While the role of muonium HFS in determining the fine structure constant was made obsolete by the highly precise $\alpha$ obtained from the measurements of the electron anomalous magnetic moment $a_e$ \cite{hfg2008} and the recoil frequency of the $^{133}Cs$ atoms \cite{pyzm2018}, it remains the best source for the precise value of the muon-electron mass ratio.

After a twenty years lull a new MuSEUM experiment on measuring the muonium HFS and the muon-electron mass ratio is now in progress at J-PARC, see, e.g., \cite{kshim2018}. The goal of the experiment is to reduce the experimental uncertainties of the muonium HFS and muon-electron mass ratio by an order of magnitude. As a byproduct the experimental team hopes to obtain limits on possible new physics contributions to muonium HFS. A proper estimate of the uncertainty of the theoretical prediction is critical in comparison between theory and experiment and figuring out the limits on new physics. Meanwhile it is now for almost twenty years two discrepant estimates of this uncertainty exist in the literature. The uncertainty in the CODATA adjustments of the fundamental physical constants \cite{codata2000,codata2002,codata2006,codata2010,codata2014} is roughly two times lower than this uncertainty in \cite{egs2001,egs2007} and some other theoretical  papers on muonium. This discrepancy was on stark display at the recent Osaka workshop on Physics of Muonium and Related Topics, see e.g.,  \cite{mie2018}.  The CODATA adjustments of the fundamental physical constants is a highly respected and reliable source, and the two times lower error bars cited in \cite{codata2000,codata2002,codata2006,codata2010,codata2014} found their way in experimental and theoretical papers on muonium HFS, too numerous to cite them here.

Below we will derive the uncertainty of the current theoretical prediction of the HFS in muonium and slightly improve its estimate in \cite{egs2001,egs2007}. This improvement is made possible by the new theoretical contributions and more accurate values of the fundamental physical constants that were obtained after the reviews \cite{egs2001,egs2007} were published. We trace out the origin of the two times lower error bars in \cite{codata2000,codata2002,codata2006,codata2010,codata2014} and explain why they cannot be used for comparison between theory and experiment.

\section{Zeeman Splitting and Experimental Measurements of Muonium HFS}

Let us describe schematically how muonium HFS and the muon-electron mass ratio were measured in the up to the present moment most precise LAMPF experiments \cite{mbb1982,lbd1999}. Measurements were done at nonzero magnetic field and two transition frequencies $\nu_{12}$ and $\nu_{34}$ between the Zeeman energy levels were measured. An elementary quantum mechanical calculation leads to the Breit-Rabi formulae for these frequencies (see, e.g., \cite{vwh1990,lbd1999})

\beq \label{zeemanspl}
\begin{split}
\nu_{12}&=-\frac{\mu_\mu B}{h}+\frac{\Delta\nu}{2}\left[(1+x)-\sqrt{1+x^2}\right],\\
\nu_{34}&=\frac{\mu_\mu B}{h}+\frac{\Delta\nu}{2}\left[(1-x)+\sqrt{1+x^2}\right],
\end{split}
\eeq

\noindent
where $x=(\mu_\mu-\mu_e)B/(h\Delta\nu)$\footnote{The minus sign in the definition of $x$, unlike the plus in \cite{mbb1982,lbd1999}, arises because we assume that $\mu_e$ is negative.} is proportional to the external magnetic field $B$. This field $B$ is calibrated  by measuring the Larmor spin-flip  frequency $h\nu_p=2\mu_pB$, where $\mu_p$ is the proton magnetic moment. We represent all magnetic moments in terms of total magnetic moments and do not write them as products of the respective Bohr magnetons and $g$-factors as in \cite{mbb1982,lbd1999,codata2014,codata2000,codata2002,codata2006,codata2010} to make the formulae more transparent. We can always restore the $g$-factors  that we swallowed in magnetic moments later if we wish.

Transition frequencies $\nu_{12}$ and $\nu_{34}$ and the spin-flip frequency $\nu_p$ were measured in the LAMPF experiments \cite{mbb1982,lbd1999}. All other parameters in \eq{zeemanspl} except the hyperfine splitting at zero field $\Delta\nu$ and the muon magnetic moment $\mu_\mu$  are known with a high accuracy. Then \eq{zeemanspl} turns into a system of two equations with two unknowns. Solving these equations we obtain

\beq
\begin{split}
\Delta\nu&=\nu_{12}+\nu_{34},\label{strangeda}\\
\frac{\mu_\mu}{\mu_p}
&=\frac{4  \nu_{12}\nu_{34} + \nu_p \frac{\mu_e}{\mu_p} (\nu_{34}-\nu_{12})}{\nu_p \left[\nu_p \frac{\mu_e}{\mu_p} -
   (\nu_{34}-\nu_{12})\right]}.
\end{split}
\eeq

\noindent
These $\Delta\nu$ and  $\mu_\mu/\mu_p$ are the experimental values of HFS at zero field and of the ratio of muon and proton magnetic moments obtained in the LAMPF experiments \cite{mbb1982,lbd1999} (we skip here all hard experimental problems).

The ratio of the electron and proton magnetic moments was measured with very high accuracy, see, e.g., \cite{codata2014}. One can use this ratio together with the LAMPF result for ${\mu_\mu}/{\mu_p}$ to obtain the ratio of the electron and muon magnetic moments. This last ratio is in its turn a product of the ratios: ratio of the electron and muon $g$-factors and ratio of their masses. The electron and muon in the muonium atom are not free, and one should remember that in this case additional quantum electrodynamic (QED) binding corrections to the $g$-factors arise (see, e.g., \cite{cdps2018} and references therein). These corrections do not exist in the case of free electron and muon, and to calculate the electron-muon mass ratio we need to take them into account on par with the QED corrections to the free electron and muon $g$-factors. Both the binding corrections and QED corrections to the free $g$-factors depend on the mass ratio, see collection of all corrections e.g., in \cite{codata2014}. This dependence is accompanied by so high powers of the fine structure constant that transition from the magnetic moments to mass ratio does not introduce an additional uncertainty in the mass ratio we obtain in this way. Combining the full QED theory of electron and muon $g$-factors, known with high precision ${\mu_e}/{\mu_p}$, and ${\mu_\mu}/{\mu_p}$ measured in the LAMPF experiment one obtains an experimental value of electron-muon mass ratio \cite{lbd1999}. The results of the two LAMPF experiments were summarized in \cite{mbb1982,lbd1999}

\beq \label{exphfslap}
\Delta \nu^{ex}_{\scriptscriptstyle HFS}(Mu)=4~463~302~776~(51)~\mbox{Hz}, \quad \delta=1.1\times 10^{-8},
\eeq
\beq \label{massratioexp}
\left(\frac{m_\mu}{m_e}\right)_{ex}=206.768~277~(24),\quad \delta=1.2\times10^{-7}.
\eeq

\section{Theoretical Prediction of Muonium HFS and its Uncertainty}

Theoretical QED formula for HFS in muonium has the form

\beq \label{formtheqed}
\Delta \nu_{\scriptscriptstyle HFS}=\nu_{\scriptscriptstyle F}\left[1+F\left(\alpha,Z\alpha,\frac{m_e}{m_\mu}\right)\right]+\Delta\nu_{\scriptscriptstyle weak}+\Delta\nu_{th},
\eeq

\noindent
where the Fermi frequency is

\beq \label{fermifr}
\nu_{\scriptscriptstyle F}=\frac{16}{3}Z^4\alpha^2 \frac{m_e}{m_\mu}
\left(\frac{m_r}{m_e}\right)^{3}c\:R_{\infty},
\eeq

\noindent
$R_\infty$ is the Rydberg constant, $c$ is the speed of light, $Z=1$ is the muon charge in terms of the positron charge, $m_r=m_em_\mu/(m_e+m_\mu)$ is the reduced mass,  function $F(\alpha,Z\alpha,m_e/m_\mu)$ is a sum of all known QED contributions, $\Delta \nu_{\scriptscriptstyle weak}$ is the weak interaction contribution, and $\Delta\nu_{th}$ is the estimate of all yet uncalculated terms. Explicit expressions for all terms on the right hand side (RHS) in \eq{formtheqed} are collected in \cite{egs2001,egs2007,codata2014}.

To obtain a theoretical prediction for HFS and its uncertainty we plug the values of all constants known independently of this very theoretical formula on the RHS hand side of \eq{formtheqed}. Currently the relative uncertainty of the Rydberg constant $\delta R_\infty=5.9\times10^{-12}$ \cite{codata2014}, and the relative uncertainty of the fine structure constant $\delta_\alpha=2.3\times 10^{-10}$ \cite{codata2014}. Nothing would change in the discussion below if we would use the relative uncertainty of $\alpha$ obtained from measurements of $a_e$ \cite{hfg2008} and/or recoil frequency of $^{133}Cs$ \cite{pyzm2018}. The least precisely known constant on the RHS in \eq{formtheqed} is the experimental electron-muon mass ratio from \eq{massratioexp} that respectively  introduces the largest contribution to the uncertainty of the theoretical prediction for HFS. We also need to take into account the uncertainty $\Delta\nu_{th}$ that is due to the uncalculated contributions to the theoretical formula in \eq{formtheqed}. The estimate of this uncertainty is relatively subjective, we consider 70 Hz to be a fair estimate \cite{es2014prd89,es2014prl,es2014}. In \cite{codata2014} uncertainty due to the uncalculated terms is assumed to be 85 Hz. We will use 70 Hz as an estimate of the uncalculated terms, but our conclusions below would not change if we would adopt the estimate from \cite{codata2014}. After simple calculations we obtain the theoretical prediction for the muonium HFS\footnote{All fundamental constants used in these calculations can be found in \cite{codata2014} and/or in \cite{pdg2018}.}

\beq
\Delta \nu^{th}_{\scriptscriptstyle HFS}(Mu)=4~463~302~872~(511)~(70)~(2)~\mbox{Hz}.
\eeq

\noindent
The first uncertainty is due to the uncertainty of $(m_\mu/m_e)_{ex}$, the second one is due to the uncalculated theoretical terms, and third is due to the uncertainty of $\alpha$. This last uncertainty is too small for any practical purposes and can be safely omitted.

We see that the uncertainty of the theoretical prediction is dominated by the uncertainty of the experimental mass ratio $m_e/m_\mu$, and to reduce it one should measure the mass ratio with a higher accuracy. The second largest contribution to the uncertainty is due to the uncalculated terms in the theoretical formula for HFS. Combining uncertainties we obtain

\beq \label{theprhfsper}
\Delta \nu^{th}_{\scriptscriptstyle HFS}(Mu)=4~463~302~872~(515)~\mbox{Hz},\quad \delta=1.2\times 10^{-7}.
\eeq

\noindent
We can compare this theoretical prediction for HFS with the result of the experimental measurements \cite{mbb1982,lbd1999} in \eq{exphfslap}. Theory and experiment are compatible but the theoretical error bars are too large due to relatively large experimental uncertainty  of the mass ratio $(m_\mu/m_e)_{ex}$.

In this situation it is reasonable to invert the problem and use the QED theoretical formula for muonium HFS in \eq{formtheqed} and the experimental result for HFS in \eq{exphfslap} to find a more precise value of the mass ratio. We obtain

\beq
\frac{m_\mu}{m_e}=206.768~281~(2)(3),
\eeq

\noindent
where the first uncertainty is due to the uncertainty of $\Delta \nu^{ex}_{\scriptscriptstyle HFS}$ and the second uncertainty is due to uncalculated terms in $\Delta \nu^{th}_{\scriptscriptstyle HFS}(Mu)$ in \eq{formtheqed}. Combining uncertainties we obtain

\beq \label{massratiofrhfs}
\frac{m_\mu}{m_e}=206.768~281~(4),\qquad \delta=2\times 10^{-8}.
\eeq

\noindent
This value of the mass ratio is compatible but an order of magnitude more accurate than the experimental mass  ratio $(m_\mu/m_e)_{ex}$ in \eq{massratioexp}. Hyperfine splitting in muonium is the best source for a precise value of the electron-muon mass ratio.

It is not by chance that the uncertainty in \eq{massratiofrhfs} practically coincides with the uncertainty of the mass ratio obtained as a result of the CODATA adjustment \cite{codata2014}. The QED formula in \eq{formtheqed} together with the experimentally measured HFS was used in the adjustment, and since the procedure described above produces by far the most precise value of the mass ratio, the result of the adjustment and its uncertainty should practically coincide with the value of the mass ratio in \eq{massratiofrhfs}.

\section{CODATA Estimate of the Theoretical Uncertainty}

\noindent
The magnitude of the theoretical prediction for muonium HFS in \eq{theprhfsper} almost exactly coincides with the respective prediction in \cite{codata2014}, while the uncertainty of this theoretical prediction in \eq{theprhfsper} is roughly two times larger than the respective uncertainty in 2014 CODATA adjustment of the fundamental physical constants (see  eq.(216) in \cite{codata2014}). Identical  uncertainty can be found in all 1998-2014 CODATA adjustments \cite{codata2000,codata2002,codata2006,codata2010} and this discrepancy should be explained.

The difference between the uncertainties of the theoretical prediction for muonium HFS in the adjustments and in this work is due exclusively to the estimate of the experimental error of the mass ratio in \eq{fermifr}. It looks as if the uncertainty of the mass ratio used in adjustments to calculate the muonium HFS and its uncertainty according to \eq{formtheqed} and \eq{fermifr} is roughly two times lower than the uncertainty of the experimental mass  ratio in \eq{massratioexp}. Let us figure out how this could happen. It can be seen from eq.(223) in \cite{codata2014} and similar equations in  \cite{codata2000,codata2002,codata2006,codata2010}. This equation (223) in \cite{codata2014} is just another form of \eq{strangeda} for the ratio of the muon and proton magnetic moments. Let us transform \eq{strangeda} to the form used in the adjustments. We notice that the product $\nu_{12}\nu_{34}$ in the numerator of the RHS in \eq{strangeda} can be identically written as

\beq
4\nu_{12}\nu_{34}=(\nu_{12}+\nu_{34})^2-(\nu_{34}-\nu_{12})^2.
\eeq

\noindent
Substituting this representation in \eq{strangeda} we obtain

\beq \label{izvrat}
\frac{\mu_\mu}{\mu_p}
=\frac{(\nu_{12}+\nu_{34})^2-(\nu_{34}-\nu_{12})^2 + \nu_p \frac{\mu_e}{\mu_p} (\nu_{34}-\nu_{12})}{\nu_p \left[\nu_p \frac{\mu_e}{\mu_p} -
   (\nu_{34}-\nu_{12})\right]}.
\eeq

\noindent
To comply with the notation in \cite{codata2014} we introduce $f_p=2\nu_p$, $\nu(f_p)=\nu_{34}-\nu_{12}$ and $\Delta\nu=\nu_{12}+\nu_{34}$. In this notation  \eq{izvrat} has the form (unlike in \cite{codata2014} magnetic moments below include all relevant QED corrections, see the discussion after \eq{zeemanspl})

\beq \label{izvratcom}
\frac{\mu_\mu}{\mu_p}
=\frac{\Delta\nu^2-\nu^2(f_p)+2f_p \frac{\mu_e}{\mu_p} \nu(f_p)}{2f_p \left[2f_p \frac{\mu_e}{\mu_p} -
   \nu(f_p)\right]},
\eeq

\noindent
and coincides with eq.(223) from the 2014 CODATA adjustment \cite{codata2014}.

Let us emphasize that \eq{izvratcom}, as well as the equivalent \eq{strangeda}, contains only the experimentally measured frequencies on the RHS. We already used \eq{strangeda} to obtain the experimental value of the mass ratio in \eq{massratioexp}. The symbol  $\Delta\nu$ on the RHS in \eq{izvratcom} is nothing but the sum of two measured frequencies and it coincides with the experimental HFS at zero field in \eq{strangeda}.  No QED theory for HFS is used in \eq{izvratcom}. As we already explained (see discussion after \eq{strangeda}) it is easy to convert the LHS of \eq{izvratcom} into the mass ratio. We will assume below that such transformation is already made.

The authors of the CODATA adjustments rejected the idea of using the experimental ratio of magnetic moments (or what is effectively the same the ratio of masses) to calculate the theoretical value of HFS arguing that this ratio depends on the experimental value of HFS and one cannot use this experimental value to obtain the theoretical prediction (see \cite{codata2000}, p. 481). This is a flawed argument, because the RHS's of \eq{strangeda} and \eq{izvrat} contain only two experimentally measured frequencies and allow us to calculate (if we trust the theory of the Zeeman effect) HFS at zero field and the magnetic moments ratio measured in the LAMPF experiments. These HFS and the magnetic moments ratio arise as two different functions of two independent experimental frequencies. The possibility to write the second of \eq{strangeda} in the form of \eq{izvrat} does not mean that it becomes a function of the experimental HFS at zero field, it remains a function of two measured frequencies. It would be a function of the experimental HFS at zero field only if it did not depend on any other combination of the measured frequencies. This is not the case in \eq{izvrat}, it depends both on the sum and difference of the measured frequencies and any function of two frequencies could be written in such form. Once again, neither of the LHS's in \eq{strangeda} are functions of one another, they both are different functions of the frequencies $\nu_{12}$ and $\nu_{34}$, and we can and should use the magnetic moment ratio from \eq{strangeda} in the QED formula \eq{formtheqed} to obtain a theoretical prediction for HFS in muonium.

In the adjustments the theoretical QED  formula for the muonium HFS from \eq{formtheqed} is plugged in the numerator on the RHS of \eq{izvratcom} \cite{btpr2018} instead of $\Delta\nu$. Then the relationship in \eq{izvratcom} turns into an equation for the mass ratio

\beq \label{selfcons}
\frac{m_e}{m_\mu}=f\left(\frac{m_e}{m_\mu}\right),
\eeq

\noindent
where the function $f$ is quadratic in the mass ratio and parametrically depends on some other constants, see \eq{izvratcom}. One can solve this equation and obtain a theoretical prediction for the mass ratio and its uncertainty based on the theoretical QED formula for HFS from \eq{formtheqed}, the Breit-Rabi formula for the Zeeman energy levels and the experimentally measured transition frequencies $\nu_{12}$ and $\nu_{34}$. This is what effectively was done in CODATA adjustments \cite{codata2000,codata2002,codata2006,codata2010,codata2014}. The theoretical prediction for the mass ratio one obtains in this way has roughly two times lower error bars than the experimental mass ratio in \eq{massratioexp} and is compatible with it. One can consider this comparison as a test of the theoretical formula for HFS splitting that was used to obtain this prediction for the mass ratio. Obviously this is not the best way to obtain the prediction for the mass ratio and test the theoretical QED formula for HFS splitting. As we have already discussed, a much more precise  value of the mass ratio may be obtained using the theoretical QED formula for HFS from \eq{formtheqed} and the experimental number for HFS from \eq{exphfslap}, as discussed in the end of the previous section.

Let us return to the discussion of the uncertainty of the theoretical prediction for muonium HFS in 1998-2014 adjustments \cite{codata2000,codata2002,codata2006,codata2010,codata2014}. There the solution of \eq{selfcons} obtained with the help of the theoretical QED formula for HFS is plugged back in this very formula \cite{btpr2018} and the obtained result together with its uncertainty is declared to be the theoretical prediction for the muonium HFS and its uncertainty. The problem with this approach is that the goal now is to compare the experimental data and QED theory for HFS, and one cannot use the value of the mass ratio obtained from \eq{selfcons} as an entry in the QED formula \eq{formtheqed}. Really, the uncertainty ascribed to this mass ratio is based on the assumption that the QED formula \eq{formtheqed} has the uncertainty that is determined by the QED theory used in its derivation, but this  is exactly the assumption we want to test comparing the QED theory and the experimental data. This is clearly circular logic, one cannot use a value of a parameter obtained with the help of a theoretical formula in this very formula with the goal to test it.  To illustrate this point let us mention that using the same logic  one could plug a more precise theoretical prediction for the mass ratio from \eq{massratiofrhfs} obtained with the help of the theoretical QED formula in this very formula  and claim that the uncertainty introduced by the mass ratio in the theoretical  prediction of HFS is effectively an order of magnitude lower. This obviously makes no sense.

\section{Conclusions}

We have shown above that the uncertainty of the current theoretical QED prediction for the muonium HFS is about 515 Hz (relative uncertainty is $1.2\times 10^{-7}$), see \eq{theprhfsper}.  By far the largest contribution to this uncertainty is due to the experimental uncertainty of the muon-electron mass ratio in \eq{massratioexp}, it exceeds the uncertainty due to the uncalculated terms in the theoretical formula by about a factor of seven. The uncertainty of the theoretical prediction for muonium HFS in \eq{theprhfsper} is roughly two times larger than in the 2014 adjustment (see eq.(216) in \cite{codata2014}) and in other 1998-2014 adjustments \cite{codata2000,codata2002,codata2006,codata2010}. All these years the  underestimation of the error bars of HFS was not practically important because there were no experimental activity on measuring muonium HFS and the muon-electron mass ratio, and the adjustments produced the value of the muon-electron mass ratio with the correct error bars. Now the situation is rapidly changing. The MuSEUM experiment \cite{kshim2018} at J-PARC  is going on and its result will be obtained in a not so far future.  It is expected that the muonium HFS and the electron-muon mass ratio will be measured with an order of magnitude higher accuracy than in the old experiments \cite{mbb1982,lbd1999}. One of the goals of the MuSEUM  experiment  is to compare the theoretical prediction for the muonium HFS with the experimental results in search of new physics. A discrepancy between theory and experiment could be interpreted as a new physics effect. The proper magnitude of the error bars of the theoretical prediction for the muonium HFS is crucial for such comparison. An underestimation of these error bars could lead to an erroneous claim of a new physics discovery. I hope that the discussion above convincingly resolves the discrepancy in the literature on the magnitude of the error bars in the theoretical prediction for the muonium HFS.

\acknowledgments

I am deeply grateful to Barry~Taylor for numerous very helpful and illuminating discussions. The original version of this paper contained an about 8 Hz inaccuracy in the estimate of the uncertainty in the first parentheses in \eq{theprhfsper}. It arose due to an improper account for the reduced mass factor in the Fermi frequency in \eq{fermifr}. I am thankful to Peter Mohr for discovering this mistake.

Of course, I am solely responsible for the contents of this paper. This work is supported by the NSF grant PHY-1724638.

\end{document}